\documentclass[twocolumn,aps,showpacs,superscriptaddress,pre]{revtex4}

\usepackage{amsmath,amssymb,graphicx}
\usepackage{bm}
\usepackage{algorithmic}
\usepackage{enumerate}
\usepackage{color}
\usepackage{stmaryrd}
\usepackage{times}

\newcommand{\iK}{{\tilde K}}
\newcommand{\tK}{K_{\rm tot}}
\newcommand{\var}{{\rm var}}
\newcommand{\cp}[1]{S_{#1}}
\newcommand{\lam}{\lambda}

\begin{document}

\title{Absence of Chaotic Size Dependence for Spin Glasses on Hierarchical Lattices}

\author{Jeffrey Gertler}
\email{jeffgertler@gmail.com}
\affiliation{Department of Physics, University of Massachusetts,
Amherst, Massachusetts 01003 USA}

\author{Jonathan Machta}
\email{machta@physics.umass.edu}
\affiliation{Department of Physics, University of Massachusetts,
Amherst, Massachusetts 01003 USA}
\affiliation{Santa Fe Institute, 1399 Hyde Park Road, Santa Fe, New Mexico
87501, USA}

\begin{abstract}
We investigate the question of whether chaotic size dependence occurs on hierarichical lattices and demonstrate that it is not present in these systems.  Our results show that the metastate for spin glasses on hierarchical lattices is simple.
\end{abstract}

\maketitle

\section{Introduction}

Chuck Newman's many contributions to statistical physics include  fundamental insights into the proper definitions of ``thermodynamic limit'' for disordered systems such as spin glasses.  He and Dan Stein elucidated the notions of {\em chaotic size dependence} (CSD) and of the {\em metastate} \cite{newman:92,newman:96a,newman:97,newman:03,StNe13}.  For systems with CSD, the usual thermodynamic limit fails to exist and one cannot define a unique thermodynamic state of the system.   Instead, the infinite volume limit must be described through the metatstate, a probability distribution over thermodynamic states (see also \cite{aizenman:90}).   For a heuristic understanding of CSD, consider correlation functions in a disordered spin system.  Specifically, consider spin correlation functions within a system of size $L$ within a large environment of size $L'$ with $L'\gg L$.    Now, imagine increasing $L'$ keeping the couplings that have already been determined at smaller sizes and the boundary conditions fixed.  How do the spin correlation functions in the system change as $L'$ increases keeping $L$ fixed?  If the usual thermodynamic limit exists these correlation functions all converges to a limit.  If CSD holds, then some correlation functions in the system fail to settle down as $L'$ increases.

A primary motivation for introducing chaotic size dependence and the metastate was to settle the question of the low-temperature behavior of the Edwards-Anderson model~\cite{edwards:75}, the Ising spin glass on finite-dimensional Euclidean lattices.  The mean field Ising spin glass or Sherrington-Kirkpatrick model was solved by Parisi~\cite{parisi:79,parisi:80}.  His solution uses the replica trick and requires {\em replica symmetry breaking} (RSB).   The RSB solution displays a infinitely many thermodynamic states and a number of other related properties that are quite counter to standard intuition about the nature of the low temperature phase of spin systems.  The question that naturally arises is whether the counterintuitive features of the RSB solution also hold for the Edwards-Anderson spin glass.   An alternative scenario holds that the mean field solution is misleading and the Edwards-Anderson model behaves more like the ferromagnetic Ising model with a simple thermodynamic limit consisting of a single pair of pure states related by a global spin flip.  This simpler picture was by developed by McMillan \cite{mcmillan:84b}, Bray and
Moore \cite{bray:86}, and Fisher and Huse
\cite{fisher:86,fisher:87,fisher:88} and goes by the name {\em droplet scaling}.  After several decades of intense study and controversy, the question of which general scenario is correct remains open.   However, Chuck and Dan's work on CSD and the metastate has radically sharpened the question and yielded insights so that at least we now know what the question is and what it would mean for something like RSB to hold for spin systems on finite-dimensional Euclidean lattices.  The resulting {\em non-standard} RSB metastate~\cite{newman:96a,read:14,BiFeMaMaMaMoPaRiRu17} has several strange features, including CSD and support on an uncountable infinity of thermodynamic states.   Chuck and Dan also introduced an alternative, perhaps more plausible, scenario also displaying CSD, called the {\em chaotic pairs} picture \cite{newman:92}.  In chaotic pairs, the support of the metastate is thermodynamic states each consisting of a single pair of pure states.  In chaotic pairs, for a given large volume, one sees only two pure state related by a spin flip whereas for non-standard RSB, one sees evidence of many pure states.  In both cases, there is CSD so that as the system size increases, the observed pure states change.

While we wait for the breakthrough that finally settles the question of the nature of the low temperature phase of the Edwards-Anderson model, it is useful to seek guidance from simpler systems.  One such simpler system is the spin glass on a hierarchical lattice.   The study of disordered spin systems on hierarchical lattices has a long history \cite{JaChWo77,southern:77,mckay:82,bray:84,gardner:84,moore:98,drossel:01, MaCa94,KiDo81,McBeKi82,BoBeCo91,ChBi17}.   Analyzing spin systems on a hierarchical lattices often yields better qualitative results than mean field theory, equivalent to spins on the complete graph.  For example, hierarchical lattices can be assigned an effective dimensionality and the behavior the system as a function of dimensionality can be studied while the complete graph is effectively infinite dimensional.  The behavior of spin systems on hierarchical lattices is usually analytically tractable or at least amenable to simple numerical simulations.  A key motivation for studying spin systems on hierarchical lattices is the Migdal-Kadanoff real space renormalization group scheme, which was shown to be equivalent to solving the spin model on a hierarchical lattice~\cite{berker:79}.   As in the case of mean field theory, the applicability of the results to Euclidean lattices must be treated with skepticism.

There have been a number of studies of the Ising spin glass on hierarchical lattices \cite{southern:77,mckay:82,bray:84,gardner:84,moore:98,drossel:01}.  Gardner \cite{gardner:84} showed that the overlap distribution for the Ising model on the diamond hierarchical lattice consists of two delta functions suggesting that for each finite size system only a single pair of thermodynamic pure states is present. A second argument against the RSB picture on hierarchical lattices is the fact that the exponent describing the dimension of  domain walls is trivially required to be $d-1$ so domains walls cannot be space filling.    Later work explains why the RSB scenario can appear to be correct for small sizes despite its absence in the infinite volume limit~\cite{moore:98}.  

Although the above observations would seem to preclude a complex metastate on hierarchical lattices, they do not obviously rule out the chaotic pairs picture \cite{newman:92}.  Here we directly confront the question of whether the metastate for the Ising spin glass on hierarchical lattices contains many pairs of pure states by studying chaotic size dependence. In the process of answering this question we must develop new recursion relations that describe the influence of the environment on pair correlations.  Our conclusion is that, at least for the two most commonly used examples of hierarchical lattices, there is no chaotic size dependence.  We thus rule out both the RSB  and the chaotic pairs scenarios for spin glasses on these hierarchical lattices.  

\section{Spin Glasses on Hierarchical Lattices}

A multigraph ${\cal G}(V,E,r)$ consists of a set of vertices $V$, edges $E$ and a function $r: E \rightarrow \{\{u,v\}: u,v \in V \mbox{ and } u\neq v \}$.  A multigraph is distinguished from a graph by the possibility of multiple edges connecting the same pair of vertices.   The Ising spin glass on a multigraph is defined by the Hamiltonian,
\begin{equation}
\label{ }
-\beta {\cal H} = \sum_{\substack{e \in E\\ r(e) = \{u,v\}}} K_{e} S_{u} S_{v}
\end{equation}
where the summation is over edges $e$ connecting vertices $u$ and $v$.  The set  $\{K_{e}\}$ consists of i.i.d. quenched random couplings on edges $e$.  Here we assume that the distribution of couplings is Gaussian with mean zero.  The variance of this Gaussian increases linearly with inverse temperature $\beta$. An Ising spin  $S_u= \pm 1$ exists on each vertex $u$.     

\begin{figure}[h]
\begin{center}
\includegraphics[scale=0.4]{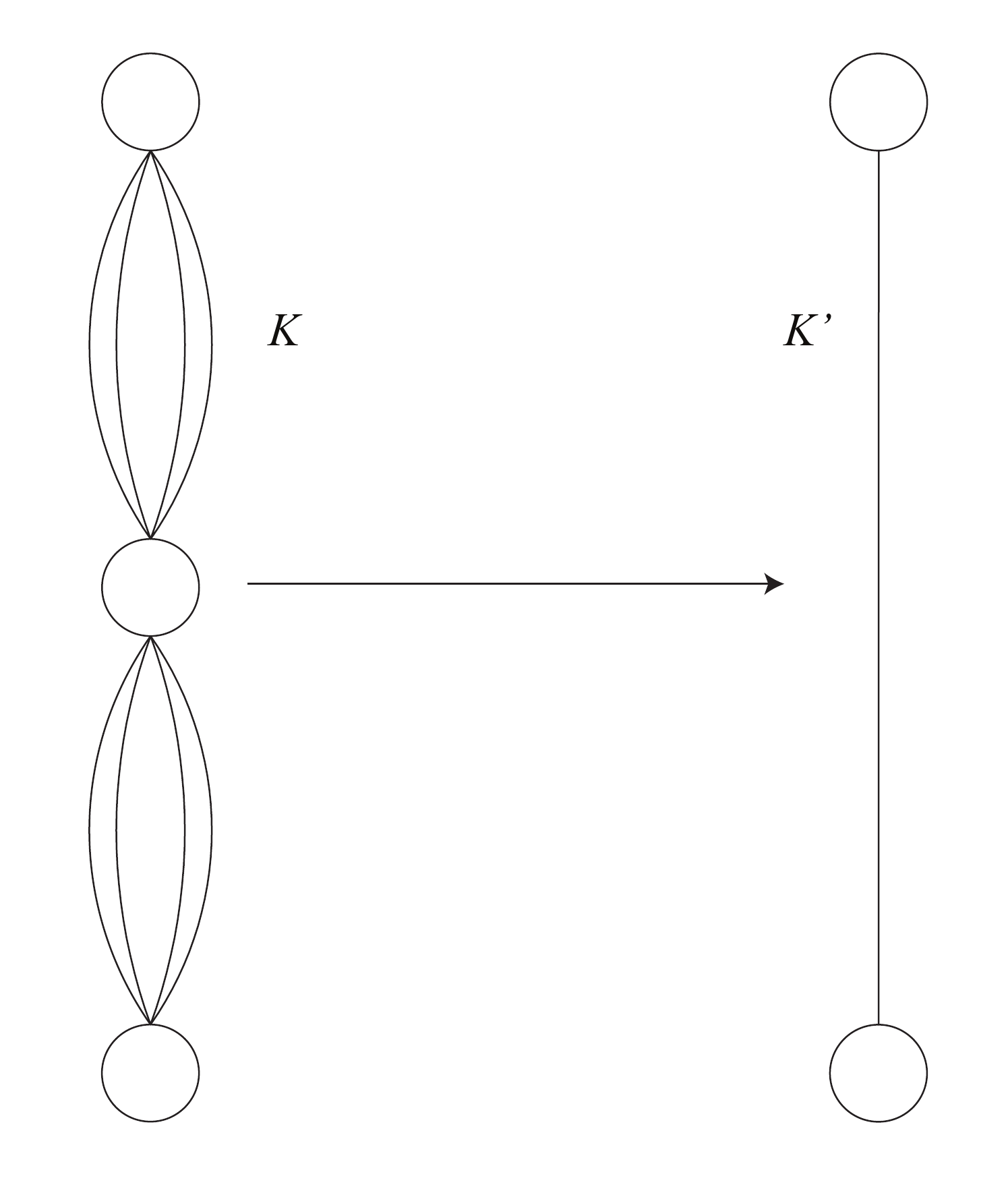}
\caption{Construction and decimation of the necklace hierarchical lattice for $b=2$ and $d=3$.  The arrow points in the direction of decimation where eight bonds $K$ are replaced by a single bond $K'$.  Construction of the hierarchical lattice proceeds in the reverse direction with the template consisting of a single bond connecting two spin replaced by eight bonds connecting three spins.
}
\label{fig:rg}
\end{center}
\end{figure}

Hierarchical lattices are multigraphs built recursively by substituting a template for each instance of an edge at the previous level of the construction.  The template and the construction process for the ``necklace" hierarchical lattice are shown in Fig.\ \ref{fig:rg}.  The necklace is parameterized by a scale factor $b$ and dimension $d$.  The number of parallel edges connecting adjacent vertices is $b^{d-1}$ while the length of the chain of vertices is $b$.  In each step of the construction, the length scale of the system increases by the factor $b$.  The figure shows the case $b=2$ and $d=3$.  

The diamond and necklace hierarchical lattices are dual to one another and the results for spin glasses are similar although some exponents differ.  
Here we focus on the necklace lattice because the analysis is slightly simpler though the qualitative results are the same for diamond lattice.

A key advantage of studying spin systems on hierarchical lattices is that real space renormalization group methods may be implemented exactly.  The decimation of spins follows the construction process of the multigraph in reverse, as shown in Fig.\  \ref{fig:rg}.   Each edge is populated with a random coupling and an effective coupling connecting the two outermost spins in the template is computed by summing over the interior spins.  This process consists of two steps.   In the first step,  all parallel couplings connecting the same pair of spins $\{u,v\}$ are added together,
\begin{equation}
\label{eq:parsum}
\iK(u,v) = \sum_{e \in r^{-1}(\{u,v\} ) } K_{e}
\end{equation}
In the second step, the linear chain of spins and couplings $\iK$ are combined into a single effective coupling $K'$, on the edge $e'$ connecting the  outermost vertices in the template, $u$ and $v$, with intermediate vertex $z$.  For the case studied here, $b=2$,  decimation results in the  relation~\cite{MaCa94},  
\begin{equation}
\label{eq:seriessum}
K'_{e'} = \frac{1}{2} \log \bigg[ \frac{\cosh(\iK(u,z) + \iK(z,v))}{\cosh(\iK(u,z) - \iK(z,v))} \bigg] .
\end{equation}
We adopt a shorthand notation for this commutative operation,
\begin{equation}
\label{ }
x \otimes y \equiv  \frac{1}{2} \log \bigg[ \frac{\cosh(x + y)}{\cosh(x - y)} \bigg] ,
\end{equation}
so that \begin{equation}
\label{ }
K'_{e'} = \iK(u,z) \otimes \iK(z,v)
\end{equation}
For spin glasses, where the couplings are chosen from a distribution, the recursion relations act on random variables so Eq.\ \ref{eq:parsum} and \ref{eq:seriessum} define a functional renormalization group for the distribution of couplings.

For the Ising spin glass on a hierarchical lattice with $d \gtrsim 2.5$ there is a zero temperature or strong disorder fixed point that controls the low temperature spin glass phase.   At the strong disorder fixed point, the coupling distribution scales under the recursion relations with a fixed form but  increasing variance.  In addition to the usual magnetic and thermal exponents, strong disorder fixed points are described by a third independent critical exponent, $\theta$, which characterizes  how the variance of the coupling distribution increases under the recursion relations according to,
\begin{equation}
\label{eq:theta}
b^{2\theta}= \var(K')/\var(K).
\end{equation}
Since the coupling between the two terminal spins of the system characterizes the stiffness of the system, $\theta$ can be identified as the stiffness exponent.

Near the strong disorder fixed point the recursion relations simplify since the variance of the coupling distribution is very large.  For random variables with very large variance the decimation operator reduces to 
\begin{equation}
\label{eq:strongtimes}
x \otimes y =  \min \big(\, |x| , |y|  \, \big)  \mbox{ sign} \big(x y \big),
\end{equation}
and, specifically, Eq.\ \ref{eq:seriessum} becomes
\begin{equation}
\label{eq:strongseries}
K'_{e'} = \min \big(\, |\iK(u,z)| , |\iK(z,v)|  \, \big)  \mbox{ sign} \big(\iK(u,z) \iK(z,v) \big).
\end{equation}

In the limit of large $d$, the recursion relations simplify further and the stiffness exponent can be evaluated analytically because the distribution of $\iK$ is Gaussian with variance given by $\sigma^2 = b^{d-1} \var(K)$.  For $b=2$, the variance of $K'$ is obtained from Eq.\ \ref{eq:strongseries} and the Gaussian form for $\iK$,
\begin{equation}
\begin{split}
\label{ }
\var(K') &=  \frac{1}{ \pi \sigma^2} \int_{-\infty}^{\infty} dx \int_{-|x|}^{|x|} dy \, y^2 e^{-(x^2+y^2)/2\sigma^2} \\
&= (\frac{\pi -2}{\pi})\sigma^2,
\end{split}
\end{equation}
so the stiffness exponent is given by 
\begin{equation}
\label{ }
2 \theta = d-1 -\log_2\bigg(\frac{\pi}{\pi-2}\bigg) \approx d- 2.46.
\end{equation}
Although this results was obtained in the large $d$ limit it remains a good approximation for small $d$ because the fixed distribution is close to Gaussian~\cite{southern:77,bouchaud:03}.  One conclusion is that the lower critical dimension on the diamond hierarchical lattice is about 2.5~\cite{boettcher:05d}. 

\section{Chaotic Size Dependence on Hierarchical Lattices}
\begin{figure}[h]
\begin{center}
\includegraphics[scale=0.6]{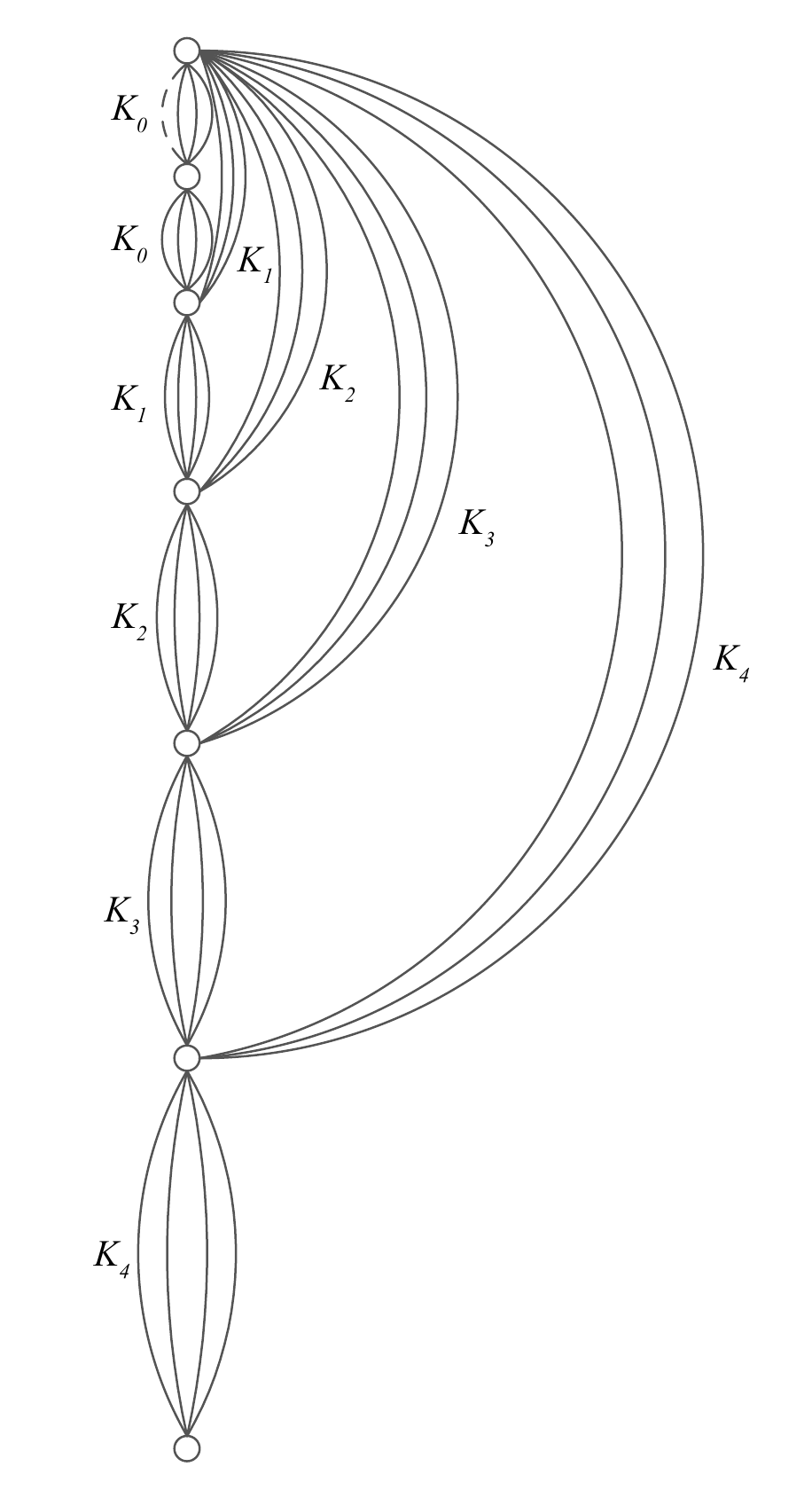}
\caption{A system, represented by a single bond $K_0$ in an environment that is $h=4$ levels larger than the system.}
\label{fig:full_diag}
\end{center}
\end{figure}

In order to investigate CSD we must understand how the environment affects a system embedded in a larger environment.  Figure \ref{fig:full_diag} shows such a situation.  In this figure, renormalization has been carried out so that the system is represented by its two terminal vertices and a single edge, shown as a dotted line. The environment is shown as extending to five levels in the hierarchy  above the level of the system.  In shorthand notation used hereafter,  $K_i$ is the random variable representing the distribution of renormalized couplings at level $i$ above the level of the system.  We suppose the environment is subject to free boundary conditions so the influence of the environment on the system is via coupling through the environment between the terminal spins of the system.  If this coupling is at least as strong as the system coupling and changes sign as the environment grows, there is CSD.  On the other hand, if this coupling converges to a limit, then CSD does not occur.

Figure \ref{fig:full_diag} shows the system located at the top end of the environment so that it includes the upper terminal spin of the environment.  The system may exist anywhere within the environment so this arrangement would seem to be a special case.  However, for an environment a given factor larger than the system, the bonds coupling the terminal vertices of the system through the environment have the same connectivity, independent of the location of the system.  This ``translation invariance'' means that we need only carry out the calculation for the system position shown in Fig.\ \ref{fig:full_diag}.   This invariance is essentially equivalent to the fact that the decimation relation ``$\otimes$" is commutative.

\begin{figure}[h]
\begin{center}
\includegraphics[scale=0.6]{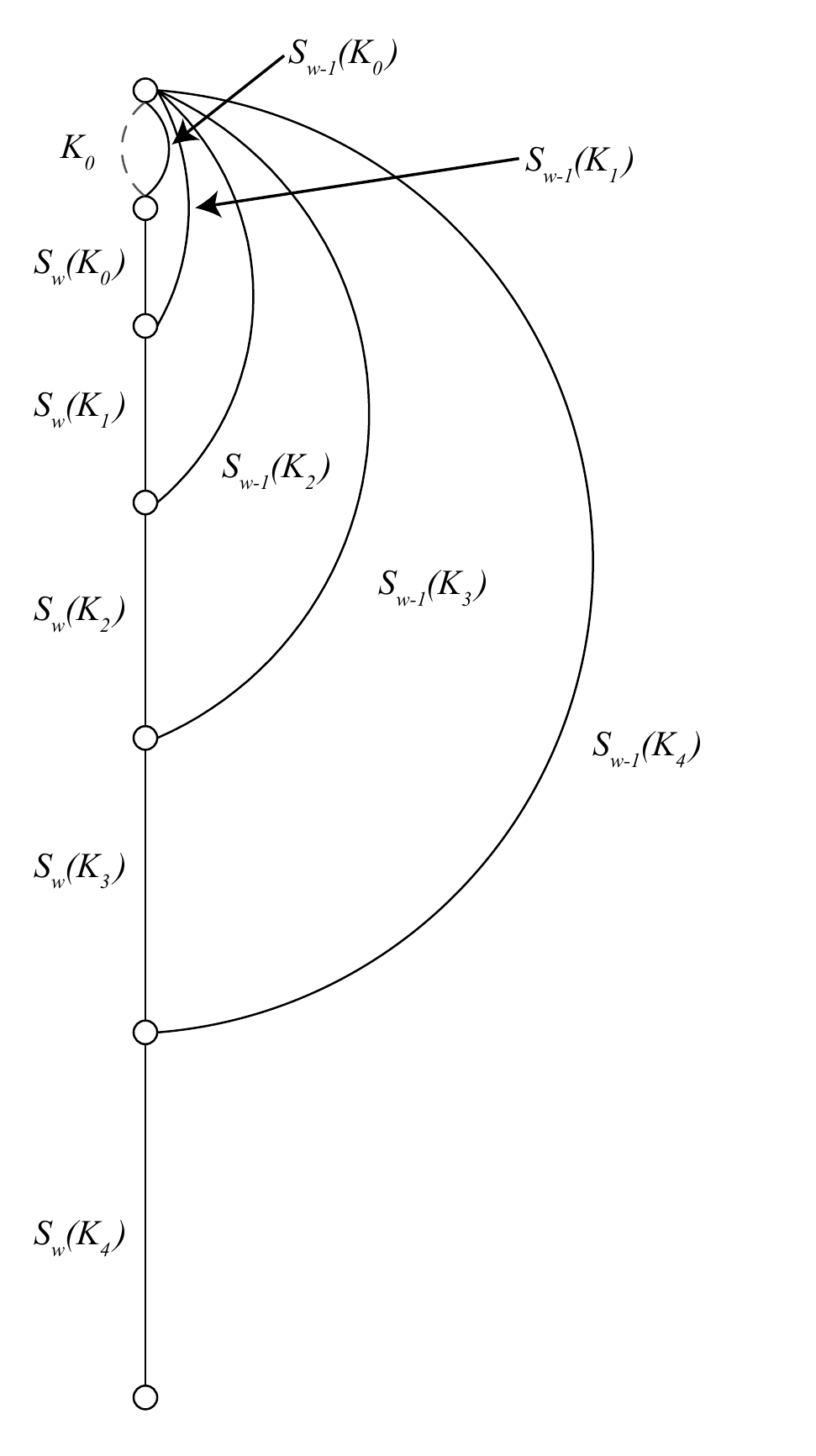}
\caption{The same system and environment as shown in Fig.\ \ref{fig:full_diag} but with parallel bonds replaced by a single bond using the notation of Eq.\ \ref{eq:cp}. }
\label{fig:reduced_diag}
\end{center}
\end{figure}

Figure \ref{fig:reduced_diag} shows the environment in reduced form where parallel edges are combined into a single edge. 
We are now ready to write down recursion relations for the environmental coupling between the terminal spins of the system.  Let $B_h$ be the random variable describing the coupling through the environment between the terminal spins of the system if the environment is $h$ levels larger than the system.  It is evident that
\begin{equation}
\label{eq:b1}
B_1 =  \sum_{j=1}^n  K_i^{(j)} = \cp{w-1} (K_0) ,
\end{equation}
where $w=b^{d-1}$ and the notation $ \cp{n} (X)$ is a shorthand for the random variable obtained by adding $n$ i.i.d. random variables,  $X^{(1)} \ldots X^{(n)}$,
\begin{equation}
\label{eq:cp}
\cp{n} (X) \equiv \sum_{j=1}^n  X^{(j)} .
\end{equation}
The recursion relations for the environmental couplings are expressed as a pattern replacement rule rather than an equation.  The replacement that produces $B_{h+1}$ from $B_h$ is
\begin{equation}
\label{eq:replace}
\cp{w-1} (K_{h-1}) \rightarrow \cp{w-1} (K_{h-1})+  \cp{w} (K_{h-1}) \otimes  \cp{w-1} (K_h ).
\end{equation} 
For example, $B_2$ is obtained from $B_1$ using the replacement rule,
\begin{equation}
\label{eq:b2}
B_2 =  \cp{w-1} (K_{0}) +  \cp{w} (K_{0}) \otimes  \cp{w-1} (K_1) .
\end{equation}
and, expanding the environment one step further, we obtain
\begin{equation}
\begin{split}
\label{eq:b3}
B_3 &=  \cp{w-1} (K_{0}) \\
&+ \cp{w} (K_{0}) \otimes \big[ \cp{w-1} (K_{1} )+ \cp{w}( K_{1}) \otimes  \cp{w-1} (K_2) \big] .
\end{split}
\end{equation}

The net coupling $\tK$ between the terminal spins of the system is given by the sum of the internal coupling $K_0$ and the environmental coupling $B_h$,\begin{equation}
\label{ }
\tK = K_0+B_h
\end{equation} 

In the low temperature phase of a  large system, the distribution of $\tK$ is the sum of two independent random variables each with mean zero and large variance so it also has mean zero and large variance.  Thus the correlation  between the terminal spins is almost always determined simply by the sign of $\tK$.  The existence of CSD is thus equivalent to $\tK$ changing sign indefinitely as $h$ increases.  It is important to note that in studying CSD, we require that the system and the environment up to level $h$ are held fixed as the environment expands and $h$ increases.  

In the low temperature phase of a  large system, the distribution of $K_i$ is close to the strong disorder fixed distribution so that, in the replacement rule that defines $B_h$, we may replace  $K_i$ by $\lam^i K_0$ where $\lam = b^\theta$ (see Eq.\ \ref{eq:theta}).  To simplify the notation, we now set $K_0$ to $K$ and, without loss of generality, normalize $K$ to have unit variance.  In the strong disorder limit, the expression for $B_2$ becomes,
\begin{equation}
\label{eq:b2s}
B_2 =  \cp{w-1} (K) +  \cp{w} (K) \otimes \lam \cp{w-1} (K) .
\end{equation}
and the pattern replacement to go from $B_h$ from $B_{h+1}$ is
\begin{equation}
\begin{split}
\label{eq:replace}
\lam^{h-1}&\cp{w-1} (K) \rightarrow \\
&\lam^{h-1}\cp{w-1} (K)+ \lam^{h-1}  \cp{w} (K) \otimes \lam^{h}  \cp{w-1} (K) .
\end{split}
\end{equation} 
Since the replacement rule in the strong disorder regime depends on $h$ only through the explicit power of $\lambda$ we can write a recursive equation for $B_h$ in terms of $B_{h-1}$,
\begin{equation}
\label{eq:b2r}
B_h=  \cp{w-1} (K) +  \cp{w} (K) \otimes \lam  B_{h-1}, 
\end{equation}
which, together with the initial condition $B_0=0$, determines the distributions of environmental couplings for all sizes $h$.

Observing that the two terms in Eq.\ \ref{eq:b2r} are independent and recalling Eq.\ \ref{eq:strongtimes} we have the following lower and upper bounds on the variance of $B_h$,
\begin{equation}
\label{ }
w-1 \leq \var(B_h) \leq 2w-1.
\end{equation}
While this bound assures us that the environmental coupling is of the same order as the internal coupling, it does not rule out CSD since a sequence of realizations of the environmental coupling may not settle down.     

Let $b_1, b_2, \ldots, b_k, \ldots$ be a sequence of realizations of environmental couplings obtained by expanding the environment so that $b_k$ is the realization of the environment at level $k$ above the system level. We stress again that $b_{k+1}$ is obtained from $b_k$ without changing any of the couplings already chosen for $b_k$.  Our main result is that this sequence almost surely becomes a constant sequence after a finite number of steps.  The result is based on ``unrolling'' the expression for $b_k$ making more and more terms explicit.  We shall find inequalities that are sufficient to establish the existence of a $k'$ such that no further changes  occur in the sequence after $k'$: $b_k=b_{k'}$ for all $k \geq k'$. 

Consider the expression for $b_k$ with $k \geq 3$, explicitly shown to three levels in the hierarchy above the system, 
\begin{equation}
\label{eq:bk}
b_k =  x^{(0)} + x^{(1)} \otimes \lam  \tilde{b}_{k,1}  ,
\end{equation}
where 
\begin{equation}
\label{eq:tildeb1}
\tilde{b}_{k,1} =  x^{(2)} + x^{(3)} \otimes \lam \tilde{b}_{k,2}.
\end{equation}
The random variates $x^{(0)}$ and $x^{(2)}$ are independently chosen from $\cp{w-1} (K)$ while  $x^{(1)}$ and $x^{(3)}$ are independent random variates chosen from $\cp{w} (K)$.  Finally, $\tilde{b}_{k,2}$ depends on $k$ and is a complicated random variate obtained from application of the replacement rules. For the present, we do not need to know anything about $\tilde{b}_{k,2}$.   
Now suppose that it is the case that the random variates appearing in Eqs.\ \ref{eq:bk} and \ref{eq:tildeb1} satisfy the inequality,
\begin{equation}
\label{eq:ineq}
\lam(|x^{(2)}| - |x^{(3)}|)  > |x^{(1)}| .
\end{equation}
Keeping in mind Eq.\ \ref{eq:strongtimes}, it is straightforward to see that if this inequality holds then the value of $\tilde{b}_{k,2}$ is irrelevant and for all $k \geq 3$,
\begin{equation}
\label{ }
b_k =x^{(0)} + x^{(1)} {\rm sign}(x^{(2)}),
\end{equation}
Another way of saying this is that the quantity,  $x^{(1)} \otimes \lam  \tilde{b}_{k,1}$, which potentially depends on $k$,  can be replaced by the constant quantity $x^{(1)} {\rm sign}(x^{(2)})$ for all $k \geq 3$ and all coupling more than two levels above the system level are irrelevant to $\tK$.
The event defined in Eq.\ \ref{eq:ineq} occurs with some probability $p>0$ and is a sufficient condition for the sequence of boundary couplings to be constant beyond level 2.  Note that in the large $d$ limit where $\lam \gg 1$, $p \rightarrow 1/2$.  

Suppose the event of Eq.\ \ref{eq:ineq} does not occur, we can continue to unroll the expression for $b_k$ until an inequality similar to Eq.\ \ref{eq:ineq} is satisfied.  The unrolling of $b_k$ is obtained from the pattern replacement rule for $\tilde{b}_{k,\ell}$, 
\begin{equation}
\label{eq:replacec}
\tilde{b}_{k,\ell}  \rightarrow   x^{(2 \ell)} + x^{(2 \ell+1)} \otimes \lam \tilde{b}_{k,\ell+1} 
\end{equation}
where $x^{(2 \ell)}$ is drawn from $\cp{w-1} (K)$  and  $x^{(2 \ell+1)}$ is drawn $\cp{w} (K)$.   The correctness of this replacement rule follows from the recursive expression, Eq.\ \ref{eq:b2r} which allows us to successively unroll $b_k$ leaving more couplings explicitly expressed while the remaining couplings are buried in  $\tilde{b}_{k,\ell}$.  Note that we could have started the unrolling of $b_k$ with the fully implicit equation, $b_k=\tilde{b}_{k,0}$.  Note also that, $\tilde{b}_{k,\ell}$ is only defined for $k > \ell$.

Now suppose that it is the case that at some stage of this unrolling, we have that
\begin{equation}
\label{eq:inineq}
\lam(|x^{(2 \ell )}| - |x^{(2 \ell+1)}|)  > |x^{(2 \ell-1)}| .
\end{equation}
It straightforward to see that if this inequality holds then the expression $x^{(2 \ell-1)} \otimes \lam  \tilde{b}_{k,\ell}$, which is potentially dependent on $k$ is, in fact, equal to the constant  $x^{(2\ell-1)} {\rm sign}(x^{(2\ell)})$ for all $k \geq \ell + 2$ and therefore, $b_k$ is constant for all  $k \geq \ell + 2$.
It is important to observe that all events of the form Eq.\ \ref{eq:inineq} (including the event of Eq.\ \ref{eq:ineq}) occur with the {\em same} probability $p$.  Thus, the probability that the sequence is not  constant up to level $\ell$ is  bounded by $(1-p)^\ell$ and  decays at least exponentially in $\ell$.  

\section{Discussion}
 
We have shown that the Ising spin glass on the necklace hierarchical lattice does not display chaotic size dependence.   Similar arguments lead to the same conclusion for the more commonly employed diamond hierarchical lattice.  Since both the replica symmetry breaking  and chaotic pairs scenarios  imply chaotic size dependence, we can conclude that, at least on hierarchical lattices, neither of these scenarios is correct.   Our results show that all correlation functions within a system are unaffected by distant changes in the environment once the environment has reached a sufficiently large size.  The convergence to the thermodynamic limit occurs exponentially in the level, $h$ of the environment above the system.  Thus, in terms of the ratio of length scales of the environment to the system, $L/L_0=b^h$ we  have power law convergence to the thermodynamic limit.  If the exponent describing this convergence is small, chaotic size dependence may be observed initially and the system may not settle down until the environment is too large to explore using numerical methods.   A forthcoming paper will discuss this size exponent quantitatively.

\begin{acknowledgments}

J.G. and J.M.~acknowledge support from the National
Science Foundation (Grant No.~DMR-1507506).  We acknowledge useful discussions with Dan Stein and Mike Moore.
\end{acknowledgments}

\bibliographystyle{apsrevtitle}
\bibliography{refs,/Users/machta/Dropbox/references}

\end{document}